\documentclass{aa}
\usepackage{latexsym, epsfig}
\usepackage{graphicx}
\usepackage{natbib}
\usepackage{txfonts}

\def\kms{kms$^{-1}$}

\begin{document}

\renewcommand{\thefootnote}{\fnsymbol{footnote}}

\title{A Method for Detection of Structure\thanks{Partly based on observations obtained at the
    Canada-France-Hawaii Telescope (CFHT) which is operated by the
    National Research Council of Canada, the Institut National des
    Science de l'Univers of the Centre National de la Recherche
    Scientifique of France, and the University of Hawaii and on
    observations obtained at the ESO/VLT, Proposal 70.C--0315.}}

\author{M.Gustafsson\inst{1}\and J.L.Lemaire\inst{2}\and D.Field\inst{1}}
\offprints{M.Gustafsson}

\institute{Department of Physics and Astronomy, University of Aarhus,
  DK-8000 Aarhus C, Denmark \\ \email{maikeng@phys.au.dk}\and  Observatoire de Paris \&
  Universit\'e de Cergy-Pontoise, LERMA \& UMR
8112 du CNRS, 92195 Meudon, France}

\date{For Main Journal, Diffuse Matter in Space. Received:  Accepted:} 

\abstract{In order to understand the evolution of molecular clouds it
  is important to identify the departures from self-similarity
  associated with the scales of self-gravity and the driving of turbulence.}{ A method is described based on structure functions for
  determining whether a region 
  of gas, such as a molecular cloud, is fractal or contains structure
  with characteristic scale sizes. }
 {Using artificial data containing structure it is shown that derivatives of
  higher order  structure functions provide a powerful way to detect
  the presence of 
  characteristic scales should any be present and to estimate the size
  of such structures. The method is  applied 
  to observations of hot H$_2$  in the Kleinman-Low nebula, north
  of the Trapezium stars in the 
  Orion Molecular Cloud, including both brightness and velocity
  data. The method is compared with other techniques such as Fourier
  transform and histogram techniques.}
{ It is found that the density structure, represented by H$_2$
  emission brightness in the K-band (2-2.5$\mu$m), 
  exhibits mean characteristic sizes of 110, 550, 1700 and
  2700AU. The velocity data show the presence of structure at
  140, 1500 and 3500AU. Compared with other techniques such as Fourier
  transform or 
  histogram, the method appears both more
  sensitive to characteristic scales and easier to interpret.}  
{}

\keywords{ISM: individual objects: OMC1 - ISM: structure - ISM: kinematics and dynamics}

\maketitle

\markboth{M. Gustafsson et al: }{}

\section{Introduction}

The velocity and density fields of molecular clouds have been
observed to show very complex structure. This property suggests that
structure is dominated by turbulence and this has been verified by
both observations and numerical simulations
\citep[e.g.][]{falgarone1990,boldyrev2002apj}; see also the extensive
reviews by \cite{elmegreen2004} and \cite{maclow2004}. Clouds dominated
by turbulence lack any 
characteristic scale and are inherently self-similar with a variety of
fractal dimensions. This has been  
observed over a broad range of scales, 0.02pc-100pc
\citep[e.g.][]{larson1981,chappell2001b,ossenkopf02}.  

Self-similarity and lack of
characteristic scale in molecular clouds must however  
break down at some dimension or range of dimensions associated with
star formation, as demonstrated in \cite{vannier} and \cite{gustafsson2006}.
Turbulence creates local clumps of gas and this process, known as
turbulent fragmentation, is considered the root cause of observed density
structure and may give rise to large density contrasts. 
Star formation begins to take place at smaller
scales in gravitationally unstable clumps.  The
distribution of sizes of these clumps is directly related to the
initial mass function (IMF) for star formation \citep{padoan2002}. The
predominantly turbulent era, characterized by lack of preferred scale, is
expected to end when gravitation becomes dominant in clumps. At this
stage, the density and velocity fields are not fractal, but
will contain characteristic scales. Hence certain scales may be
over-populated or under-populated relative to the model of a simple 
Kolmogorov-type cascade \citep{kolmogorov1941,frisch1995}. The range of
scales, at which clumps become gravitationally bound and turbulence 
looses its kinematically dominating effect, is important for both
theories of star formation and of turbulence. 

Characteristic scales may also be imposed on the medium by other
mechanisms than gravitational collapse. In particular energy may be
injected into the medium at
the largest scales by supernova explosions or at the tens of thousands
of AU scale by outflows from massive stars or from low mass stars at
scales of a few thousand AU. In order to understand the evolution of
molecular clouds it is essential to identify the presence of these energy
injection processes and the scales at which they occur.

The purpose of the present paper is to introduce a technique,
based on structure functions (SFs), to identify characteristic scales in the
interstellar medium - or indeed in any medium. The technique appears
more sensitive to characteristic scales, for example in spatial
brightness or velocity, than methods presently used. The method is
relatively easy to interpret in terms of preferred scale sizes and is
as easy to implement as existing methods.  

Experience has
proven that it is non-trivial to establish the scale(s) at which
self-similarity begins to break down and an array of techniques as
broad as possible should be made available. For example, four different
methods were used in \cite{vannier} in order to demonstrate that
self-similarity breaks down at star-forming scales of around 500 -
1000 AU in a small portion of the Orion Molecular Cloud (OMC1). The present contribution
should be seen in this light, not as a panacea for structure
identification but rather as
another tool for this purpose.

The methods currently available are those of (i)~\cite{blitz}, who introduced
a method of degrading the spatial resolution and establishing the
presence of preferred scales from histograms of the degraded 
images, a method used for example in~\cite{vannier} and~\cite{lacombe2004},
(ii) Fourier transform power spectra, used to pick out frequencies and
hence scales in images~\citep[e.g.][]{vannier}, (iii) clump 
decomposition \citep{stutzki1990}, (iv) $\Delta$-variance analysis \citep[][]{stutzki1998,
bensch2001}, (v) wavelet
transformation~\citep[e.g.][]{farge1992}.  

The non-fractal nature of molecular clouds at the star-forming scale
of $\sim$~500-1000 AU was first detected in \cite{vannier} for
Orion. Essentially the same preferred scales between  
500 and 1100 AU are also reported in \cite{lacombe2004}. Non-fractal nature has
also been reported at larger scales in \cite{blitz} who found a
preferred scale of 0.25-0.5 pc ~(5-10$\times 10^4\,$AU) in 
Taurus. Material may however be self-similar down to a few x
10$^{-4}$pc ($<$100 AU) in diffuse material, with clumps of
mass~$\sim$~10$^{-7}$M$_{\odot}$, as found in the photodissociation
region around the B-star HD37903 in NGC2023 \citep{rouan1997}. Here we
show that structure functions (SFs) or local derivatives of SFs
provide a powerful way to detect preferred 
scales in a medium. The method is tested on
observations of H$_2$ emission in OMC1.

\section{Method}
\label{sec:method}

The SFs of order p of a spatially resolved parameter A
are defined as 
\begin{equation}
S_p(r)=\langle |A(r')-A(r'-\delta)|^p \rangle = \langle |\Delta A|^p \rangle
\label{eq:struc}
\end{equation}
where the average is taken over all map positions $r'$ and all
distances $\delta$ such that $|\delta|=r$. The parameter A can for example
be velocity or brightness. The SFs are measures of the
spatial correlations of the parameter A. A completely random
distribution would result in a constant structure function as a
function of $r$, whereas for example a gradient in the image would
lead to a structure function increasing with r.

Structure functions have traditionally been used in turbulence theory
to quantify the properties of the turbulence.
In a predominantly turbulent medium where the velocity and density
fields are self-similar or fractal, the SFs are well
described by power laws, $S_p(r) \propto r^{\zeta(p)}$ in the
  inertial range between the driving and dissipation scale. This is found
from both theory \citep{kolmogorov1941,boldyrev2002prl} and
simulations \citep[e.g.][]{boldyrev2002apj,kritsuk2004} and simply conveys that
in turbulent regions, the velocity field, say, is spatially correlated in
such a way that larger separations correspond to larger velocity
differences on average. This is in fact a manifestation of the turbulent energy
cascade originally proposed by Kolmogorov.

On the other hand, if the region in question is not self-similar, but
contains structure with a preferred scale, the SFs
will deviate from power-laws. It is shown here that these deviations are a sensitive test of the presence of structure and may be used to reveal the size of structures. 
If the region contains only a single
structure in the map of 
parameter A of a certain size, R, it is evident that correlations will
only persist up to scale R. In the structure function this will show
up as an increase of $S_p(r)$ at scales $<$ R and a constant value at
scales $>$ R. The same behaviour will be found if there is a large number of
structures with the same size randomly distributed in the map. 
When clumps with two or more distinct preferred sizes are present, which is
most likely the case in star forming regions, the structure function
will show this effect for all sizes involved. The effect on the SFs will become successively less pronounced at any
specific size as more scales are identified. In a real system,
structures will most likely tend to group around a preferred scale with some
characteristic deviation from a mean value. This aspect turns out to
be more of a gloss on the present description than a fundamental point
but can lead to systematic errors in structure size
determination. This is described in Sect.~\ref{errors}. 

To test the ideas set out above, three maps have been constructed with 
i) a single circular clump with a
diameter of 40 pixels and an intensity distribution given by the
paraboloid function 
\begin{equation}
f(x,y)=C(1-\frac{(x-x_0)^2+(y-y_0)^2}{r^2})
\end{equation}
where $r$ is the radius of the clump, $(x_0,y_0)$ is the centre
position and C is the peak intensity,
ii) 100 
randomly distributed clumps with the same characteristics as in i) and 
iii) 100 clumps with a
diameter of 20 pixels and 20 clumps with a diameter of 100 pixels where
again all 
clumps were randomly distributed and of paraboloid
shape. In all cases the clumps have been distributed on a square 1024
by 1024 grid. Similar maps were also constructed
using conic shapes
($f(x,y)=C(1-\frac{\sqrt{(x-x_0)^2+(y-y_0)^2}}{r}$)) for the intensity
distribution in the clumps.

\begin{figure}
\resizebox{\hsize}{!}{\includegraphics{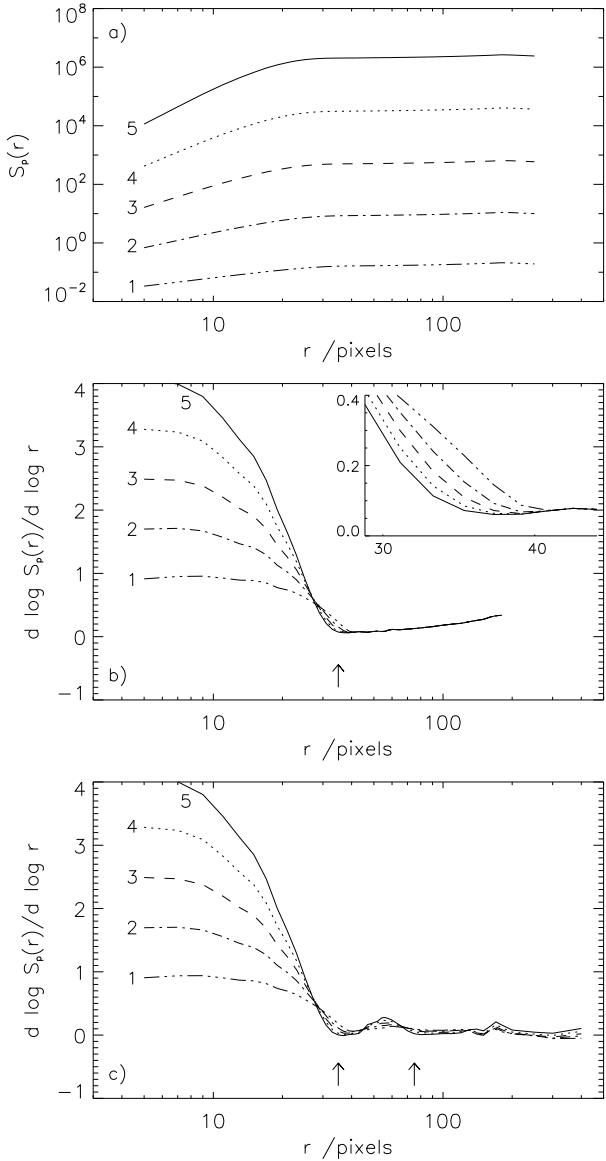}}
\caption{\small a) Structure functions of order 1-5 (labelled
  accordingly) for a 
  simulated map, case (i) (see text). b) Local logarithmic derivatives
  of the SFs shown in a) Inset: Blow-up of r=29-45
  pixels. c) Local logarithmic 
  derivatives of SFs of order 1-5 derived from the
  simulated map of case (ii).}
\label{fig:sim1}
\end{figure}

The SFs of order 1-5 for case (i) are shown in
Fig. \ref{fig:sim1}a. The SFs become roughly
constant at $r \geq 30$ pixels close to the expected value of 40. The
SFs for case  
(ii) are essentially the same. They are not shown here. The
effects on the SFs caused by the presence of clumps are more
clearly seen at higher orders, p.

Preferred
scale sizes in a map or an image result in changes in the logarithmic
slope of SFs. Therefore the local logarithmic
derivatives, $d \log S_p(r)/d \log r$, highlight the deviations
from fractality and thus we propose to use these derivatives to detect such 
scales. As seen above the presence of preferred scales causes a
decrease in  
the slope of the SFs making it roughly constant near the relevant size. This
results in a plateau in the logarithmic derivatives if larger scales
are not present, or a minimum if such scales are present. The former
case is demonstrated in Fig.~\ref{fig:sim1}b for case (i) above, that
is, for a single clump, and in Fig.~\ref{fig:sim1}c for case (ii), that
is, for 100 such clumps. 
A minimum is seen at $\sim$ 35 
pixels in both Figs.~\ref{fig:sim1}b and c, close to the imposed
structure size of 40 pixels (see 
above). The inset to Fig.~\ref{fig:sim1}b displays the derivatives around
r$\sim$ 35 in detail and shows that the position of the minimum moves
towards a lower value of r when p is increased. This occurs because 
higher orders of the structure function give more weight to higher
values of $\Delta A$ (see Eq.~(\ref{eq:struc})), thus depressing the
outer regions of clumps, which is  
the location of low values of $\Delta A$ in these simulations. Thus estimates
of scale sizes from high order SFs represent lower
bounds to the scale sizes and the estimated values can be
10-20\% too low. In Fig.~\ref{fig:sim1}c, in addition to
the first plateau or weak minimum at $\sim$~35 pixels,  
there is a secondary
plateau at $\sim$ 75 pixels, indicating the presence of a scale size
approximately twice as large as the clumps in the simulation. This is most likely due to overlapping of the randomly placed
individual clumps. By implication this method cannot distinguish between
one large clump or several smaller clumps nearly coinciding in the
same line-of-sight.

\begin{figure}
\resizebox{\hsize}{!}{\includegraphics{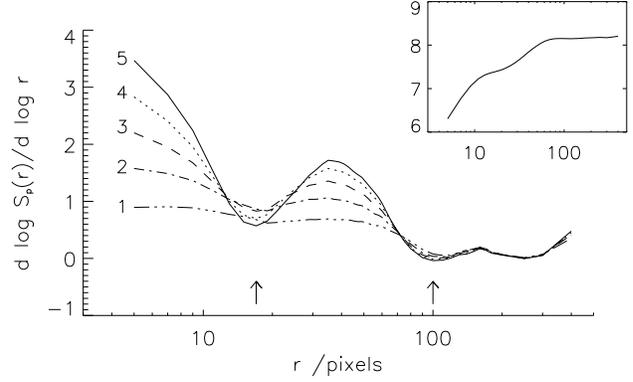}}
\caption{Local logarithmic derivatives of SFs of order
  1-5 from the simulated map of case (iii) (see text). Inset: the
  logarithm of the 5th
  order structure function of case (iii).
}
\label{fig:sim2}
\end{figure}

The local logarithmic derivatives for p=1-5 are shown in
Fig. \ref{fig:sim2} for case (iii) involving 100 clumps of diameter 20
pixels and 20 clumps of diameter 100 pixels. Here a local minimum is seen
at $\sim$ 17 pixels and $\sim$ 100 pixels, representing the 
scales of 20 and 100 pixels included in this test image. 
Note that two scale sizes will only show clearly if the scales are
well separated. If they are too close the larger scale will dominate
and suppress the smaller. In real data this means that a detected
scale size might conceal other smaller scales (see also Sect.~\ref{errors}).
The inset to Fig.~\ref{fig:sim2} shows the
corresponding structure function for 5th order, showing plateaux
at $\sim$ 15 pixels and at  $> \sim$80 pixels.
 
The same tests were performed on the maps of clumps with conical
shaped intensity distributions
and yielded similar results to those above. This suggests that effects seen
in the structure 
functions are independent of the shape of the clumps. This is of
course only valid for clump geometries which can at least crudely be
characterized by a single dimension. Tests were also carried out on
clumps with an ellipsoid shape with randomly oriented axes. These are
characterized by two sizes, 
that is, those of the major and minor axes and the results are
very similar to those of clumps with two different sets of diameters,
that is case (iii) above.

An additional property is that in the presence of real structure the
derivatives of structure 
functions of all orders tend to congregate  around the
same value at 
a local minimum or plateau. This provides a further characteristic
which aids the detection of structure. In numerical simulations of
fully developed turbulence, 
\cite{biferale2004} noted a similar tendency for the local derivatives
to accumulate at a certain value.

A further point in assigning the presence of structure is the question
of how large a
structure may be detected for a given size of image, so called
"edge-effects". In case (iii), in Fig. 2, there is already 
evidence for apparent structure at $\sim$~250 pixels which is not in
the data.  
Edge-effects have been studied here
using fractional Brownian motion structures (fBm-fractals). We have
calculated SFs and logarithmic derivatives for a
number of such fractal images. For these, the SFs are expected
to be pure power laws and the derivatives should be independent of
r. The value of the logarithmic derivative depends on the
choice of power spectral index and the order of the structure function
involved, being proportional to the latter. In passing we note that
fractal systems cannot therefore yield logarithmic derivatives of SFs
which congregate at a particular value.  

\begin{figure}
\resizebox{\hsize}{!}{\includegraphics{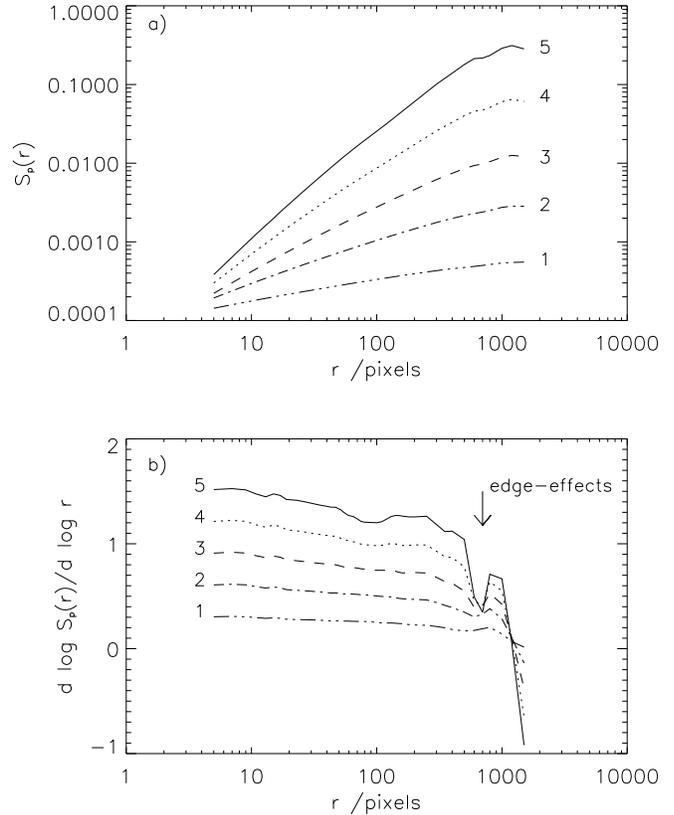}}
\caption{\small a) Structure functions of order 1-5 (labelled)
of fractal image. b) Corresponding local logarithmic derivatives of structure
functions of order 1-5.}
\label{fig:fractal}
\end{figure}

An example of these test calculations is seen in
Fig.~\ref{fig:fractal} where SFs of order 1-5 and
logarithmic derivatives for a fractal image of size 2048$\times$2048
pixels are shown. The SFs may be approximated by
power laws save at the largest scales. The logarithmic
derivatives are nearly constant up to r $\sim$ 700 pixels, any
deviations from constancy below this value being due to the imperfect
fractal nature of the simulations arising from pixelation. At around
700 pixels, a 
local minimum is found and the derivatives for p=3-5 congregate around
the value of 0.4, apparently indicating a preferred structure size. The same
analysis has been carried out on fractal images of size
1024$\times$1024, 512$\times$512 and 256$\times$256 pixels. In all
images apparent 
structure at scales of $\sim$ 1/5 - 1/3 of
the size of the map was found. Hence increasing the size of the map
also increased the absolute value of the apparent structure size. Thus
we conclude that the apparent  
structure in the fractal images is an edge-effect and that detection of scale
sizes much greater than 1/5 of the size of the map should not be
attempted as such scales cannot be reliably identified.

\section{Real data for test purposes}
\label{sec:data}
In order to illustrate the method of structure analysis described here
with real data, we use results for the emission from vibrationally
excited H$_2$ in OMC1, where excitation occurs largely through shock
excitation \citep[with a small photon excitation component: see
e.g.][]{kristensen2003}. The relevant data are
reported in~\cite{gustafsson2003}, \cite{lacombe2004}, \cite{nissen2005} and
\cite{kristensen2006}. These three datasets  
refer only to the hot component of the gas, above $\sim$~1000K, but
nevertheless form valuable datasets for the identification of
structure since they probe very small scales, from as low as $55\,$AU
\citep{lacombe2004}, and also involve the major constituent of the gas,
albeit in an excited state.  

Data in \cite{gustafsson2003} were 
obtained at the CFHT with a Fabry-Perot interferometer in 
conjunction with adaptive optics  \citep[GriF,][]{clenet}
and contain both brightness and velocity
information. Observations were performed
in the NIR K-band by scanning the $v$=1-0 S(1) H$_2$ emission line at
2.121$\mu$m. The field of view was 36\arcsec $\times$ 36\arcsec and
the pixel scale is 0\farcs035 (1024$\times$1024 pixel detector). The
dataset consists of four 
spatial and velocity resolved images, which are amalgamated into one
field of 89\arcsec$\times$67\arcsec or 0.2$\times$0.15 pc (4.1~$\cdot~10^4$ $\times$ 3.1~$\cdot~10^4$ AU) given a
distance to Orion of 460 pc \citep{bally}. The field is centred
approximately on the Becklin-Neugebauer (BN) object
(05$^{\mathrm{h}}$35$^{\mathrm{m}}$14\fs1, -05\degr22\arcmin 
22\farcs9). The spatial resolution is
0\farcs15 (70 AU) and relative velocities between adjacent resolution
elements are determined to 0.3\kms (1$\sigma$) in bright regions to
8.4\kms in the weakest regions considered here. A detailed
description of data acquisition and reduction may be found
in~\cite{gustafsson2003} with further discussion in
\cite{gustafsson2006}. We will refer to this dataset as the CFHT/GriF data.

Data in \cite{kristensen2006} were also obtained at the CFHT, with
the PUEO adaptive optics and the KIR infrared detector but without
velocity information and covering a more restricted region than the
CFHT/GriF data. Emission from the H$_2$ 1-0  
S(1) line was isolated using a narrow-band filter
($\lambda$/100). Continuum was recorded with a filter centred at
2.183$\mu$m and subtracted from the data obtained in the S(1) filter,
data being subjected to standard reduction procedures
\citep{kristensen2006}. The dataset used here is that designated as W
in \cite{kristensen2006} centered $\sim$ 18\arcsec south-west of
BN. The 
field is 36\arcsec $\times$ 36\arcsec, the  
pixel scale is 0\farcs035 and the resolution is 0\farcs45 (200AU). This
dataset will be referred to as CFHT/PUEO-KIR.

Data in \cite{lacombe2004} were obtained with VLT/NACO. The H$_2$ 1-0
S(1) line was isolated using a narrow-band filter, with continuum
subtraction as described in ~\cite{lacombe2004}. The observations used here
are those designated as the east-south-east (ESE) field,
27\arcsec$\times$27\arcsec with a pixel scale of 0\farcs027 and a
resolution of 0\farcs12 (55AU). This corresponds to the field analysed
in \cite{vannier} and 
is referred to here as the VLT/NACO data. This field lies to the
east of the CFHT/PUEO-KIR field. 

In the succeeding sections we use the new technique employing SFs for
the various datasets as follows. First we use brightness data
extracted from the CFHT/GriF. These data
cover the greatest physical extent of any of our datasets. At this
stage we ignore the velocity 
information in these data. We then consider the 
CFHT/PUEO-KIR and VLT/NACO data. In the latter
two cases we compare our results with those of the histogram method
of~\cite{blitz} and the Fourier transform method. In a third section
we use the velocity data contained in the CFHT/GriF results.

\section{Identification of Structure in OMC1} 
\label{sec:results}
\subsection{Use of the method with CFHT/GriF H$_2$ brightness data}
\label{sec:new}

\begin{figure}[!]
\resizebox{\hsize}{!}{\includegraphics{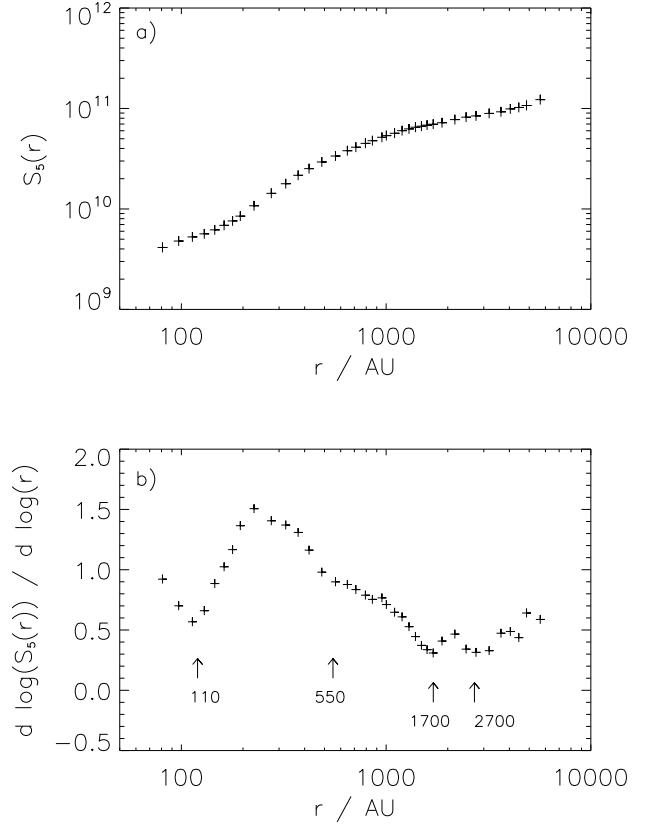}}
\caption{\small a) The 5th order structure function of velocity
  integrated brightness in OMC1 from GriF data from CFHT. b) Corresponding local logarithmic 
derivatives of the structure 
function of order 5.}
\label{fig:intensity}
\end{figure}

We use the CFHT/GriF observations described in Sect.~\ref{sec:data} to
show that 
the molecular gas in OMC1 contains a
number of preferred scales, illustrating the present technique. 
We have calculated SFs and logarithmic derivatives
for these data. Note that brightness is roughly representative of the
density structure in the excited H$_2$. 
In Fig. \ref{fig:intensity}
the structure 
function of order 5 for the
brightness and the corresponding derivatives are
shown. 
Relative errors on
values in the SFs are found using the law of
propagation of errors for uncorrelated variables. Using the notation
of Eq.~(\ref{eq:struc}), the
variance is calculated as: 
\begin{eqnarray}
\sigma^2(S_p(r))=&\left(p^2/N^2 \right)&\sum \left(
  \sigma^2(A(r'))+\sigma^2(A(r'-\delta))\right)\nonumber\\
&\times& \left|A(r')-A(r'-\delta)\right|^{2(p-1)}
\label{eq:error}
\end{eqnarray}
where the summation is performed over all pairs of pixels in the map
that satisfy $|\delta|=r $, N is the total number of such pairs and
$\sigma^2(A(r'))$ is the variance of $A(r')$.
Due to the large number of pixel pairs in our data, 2$\cdot$
10$^7$ to 4$\cdot$10$^9$, the variances are very small. Typical values of the
relative errors, $\sigma(S_p(r))/ S_p(r)$, are 10$^{-5}$ in the 2nd
order SF, 10$^{-4}$ in the 5th order SF and 10$^{-3}$ in the 10th order SF. 
Thus the errors are
negligible in SFs of order up to 10, as well as in the
logarithmic derivatives of these functions.

Local minima in the
derivatives as 
a function of r are present at r $\sim$ 110AU and r $\sim$ 1700AU,
providing clear evidence that the brightness structure, and by
implication the density distribution, in OMC1 shows preferred scales. The positions of the minima indicate that the (largely) shocked regions in the field have two preferred sizes close to 110 and 1700AU. Two less pronounced
minima or points of inflection are present at  r $\sim$ 550AU and 
r $\sim$ 2700AU, which suggests that clumping at these scales is also
present in the field, but to a lesser degree. 
As noted earlier these sizes may be representations of diameters in
near-circular clumps or major and minor axes in elliptically shaped clumps.

\begin{figure}[!]
\resizebox{\hsize}{!}{\includegraphics{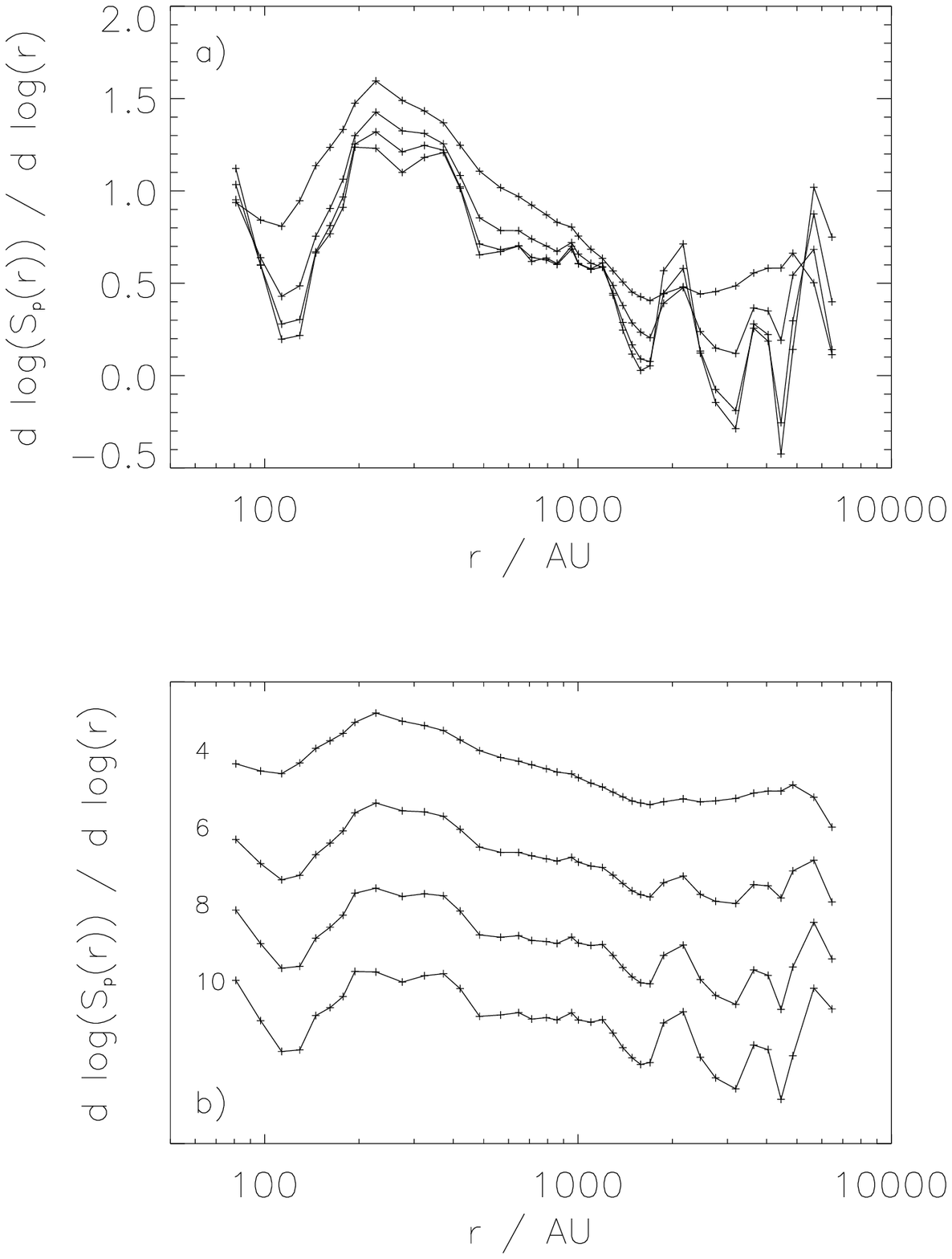}}
\caption{\small a) Local logarithmic 
derivatives of structure 
functions of order 4,6,8,10. b) As in a) but with the graphs
displaced to get a better view of the evolution.}
\label{fig:intensity1}
\end{figure}

In
Fig.~\ref{fig:intensity1}a the logarithmic derivatives of SFs of order
4,6,8,10 are displayed for the same data. This illustrates the
tendency for the derivatives of higher order SFs to congregate at the
same value, a property mentioned in Sect.~\ref{sec:method}. In
Fig.~\ref{fig:intensity1}b the derivatives are 
displaced on the ordinate to give a clearer view of the
evolution with regard to the order of the SF. As the order of the
structure function increases, the derivatives 
show more detail. 
High order SFs accentuate the presence of weaker
structure hidden in lower order SFs, resulting in more minima in
the derivatives. Thus in the limit of very high order, so far as this can be achieved within the errors of observation, derivatives will essentially pick out
the size of every clump in the region. This property is useful since
the order of the 
structure function at which a structure becomes apparent is a
measure of the prevalence of that structure. A structure that is
clearly seen in the 2nd order SF is more dominant in the region than a
structure which is first seen in the 6th order SF, say. Examples are the
scale sizes at r$\sim$ 550AU and 2700AU which are barely visible in the
derivatives of the 4th order SF, but are clearly seen in the
derivatives of the 6th order SF. These are less dominant than the scale
sizes at r$\sim$ 110AU, 1700AU, which are clearly apparent in the
derivatives of the 4th order SF.
In observational data, random errors
in the calculated values 
will however increase when higher orders are computed, thus
limiting the maximum value of p that may be used. Thus uncertainties
in observational 
data may generally be a limiting factor in this form of data analysis.

\subsection{Comparison with other methods using CFHT/PUEO-KIR and VLT/NACO data} 
\label{sec:comparison}

\begin{figure}
\resizebox{\hsize}{!}{\includegraphics{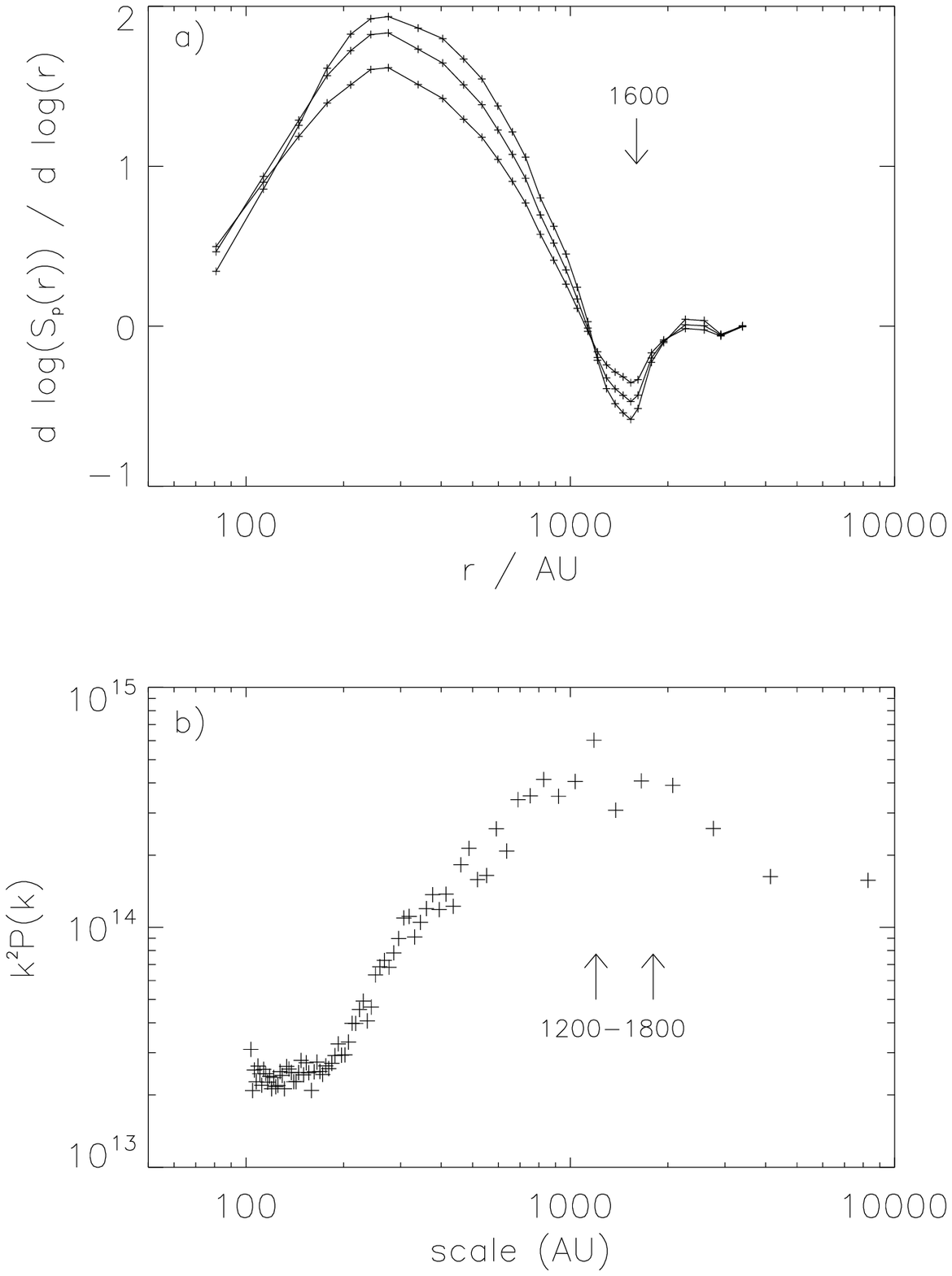}}
\caption{\small a) Local logarithmic derivatives of structure
  functions of order 4,5,6 of a region South-West of BN in OMC1
  calculated using CFHT/PUEO-KIR data. b) Fourier transform of the same set
  of data: the power spectrum multiplied by $k^2$ vs. the scale size.}
\label{fig:fourier}
\end{figure}

We first present a comparison of the method described here with the Fourier transform technique conducted on the CFHT/PUEO-KIR data of
~\cite{kristensen2006}. Figure~\ref{fig:fourier}a shows the 
derivatives of the SFs of order 4,5,6 of that region indicating a
preferred scale of 1600 AU. We now compare this with the standard
two-dimensional Fourier transform technique commonly employed to
identify scale size in images. This yields a power spectrum P(k) which
in the case of self-similar structure should be a power law in
k (k $\propto$ r$^{-1}$). Deviations from a power law are nicely illustrated by showing
k$^2$P(k) vs k, as described in
\cite{vannier}. Figure~\ref{fig:fourier}b displays 
$k^2$P($k$) vs. scale size (k$^{-1}$).  Figure~\ref{fig:fourier}b shows
that there is a range of 
scales which are over-populated. This over-population peaks at $\sim$
1200-1800 AU. The above results show good agreement between the scales
obtained from the power
spectrum and the SF derivatives, at any rate in this simple case in
which only a single scale, or a single range of scales, is present in the region.

We now turn to a comparison with a more complicated example in which
more than one scale is present. Preferred scales in OMC1 have
previously been identified by \cite{vannier} 
and \cite{lacombe2004} using a variety of techniques. We now use the
SF technique to obtain scale sizes in the VLT/NACO data. These may be
compared with scale sizes reported in \cite{lacombe2004} and
\cite{vannier}.  
Using an 
area-perimeter analysis and a Fourier transform technique \cite{vannier}
found a preferred scale around 600AU, but spanning a large and rather
poorly determined range. \cite{vannier} also used the
brightness histogram technique of \cite{blitz}, finding a preferred scale of $\sim$900 AU. \cite{lacombe2004}, with the VLT/NACO data, 	
used only the 
histogram technique finding a preferred size of 1100$\pm$130AU. 

\begin{figure}
\resizebox{\hsize}{!}{\includegraphics{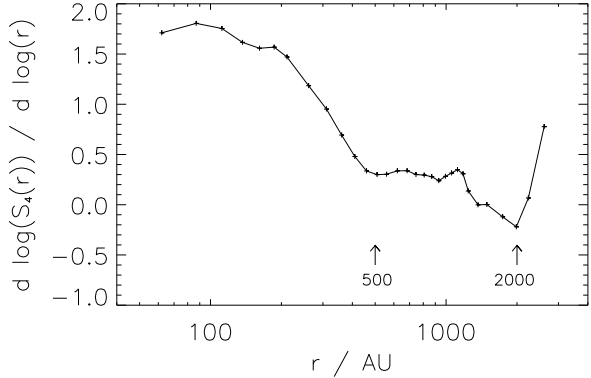}}
\caption{\small Local logarithmic derivatives of the 4th order structure
function of the ESE region analysed in \cite{lacombe2004} from
VLT/NACO data,
comparable to the region analysed in \cite{vannier}.}
\label{fig:naco}
\end{figure}

The local derivatives of the 4th order SF for the VLT/NACO data are
shown 
in Fig.~\ref{fig:naco} where two scale sizes of r$\sim$500AU and
2000AU are evident. The lower of these scales is consistent with the
findings in \cite{vannier}.
As a further comparison, we created a
two-dimensional Fourier transform to yield the power spectrum, P($k$)
of the same 
field using the VLT/NACO rather than the 3.6m La Silla data of
\cite{vannier}. The power spectrum was - as it should be - very
similar to that reported in \cite{vannier} showing marked departure
from a power law in  
$k$. Data in Fig.~\ref{fig:naco} show
that the SF method is capable of much more precision in the
identification of structure scale than a simple Fourier transform. 

In summary, there is reasonably good correspondence between the
standard technique of 2D Fourier transform and the SF method, so far
as comparison 
can be made. Both show clear departure from self-similarity and a
similar scale size, where this can be identified as in
Figs.~\ref{fig:fourier}a and b. There seems to be some discrepancy between
these two techniques and the histogram method, on the basis of work
reported in \cite{vannier} and \cite{lacombe2004}

\subsection{Use of the present method with CFHT/GriF H$_2$ velocity data}
\begin{figure}
\resizebox{\hsize}{!}{\includegraphics{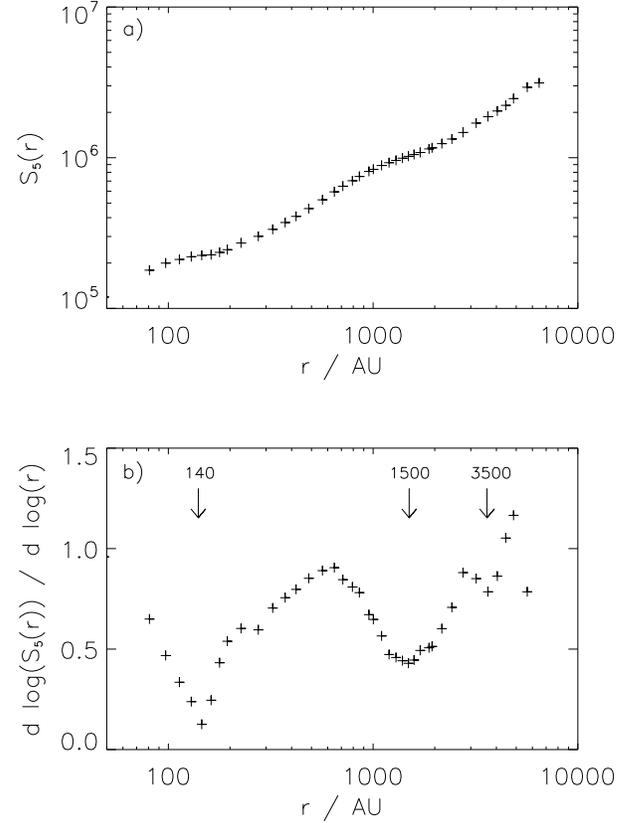}}
\caption{\small a) The 5th order structure function
of velocities in OMC1 from CFHT/GriF. b) Corresponding local
logarithmic derivatives of the 5th order structure
function.}
\label{fig:velocity}
\end{figure}

We now return to the dataset covered by the CFHT/GriF observations, as
described in Sect.~\ref{sec:data}. The 5th order structure function of
the velocity data from GriF and the corresponding derivatives are
shown in Fig.~\ref{fig:velocity}. Because the accuracy of the velocity
in each pixel depends on the brightness in the pixel (see
Sect.~\ref{sec:data}) each 
velocity difference in Eq.~(\ref{eq:struc}) has been weighted by the
product of the 
brightness, B, in the two pixels in question, thus giving more weight to
pixels with high brightness. The weighted SFs are thus
\begin{equation}
S_p(r)=\langle B(r')B(r'-\delta) |v(r')-v(r'-\delta)|^p \rangle
\label{eq:weigh}
\end{equation}
Relative errors on the calculated
values in the weighted SFs are typically 10$^{-3}$ for fifth order (p=5) and
therefore negligible. 

Figure~\ref{fig:velocity} shows that local minima in the derivatives as
a function of r are present at r $\sim$ 140AU, 1500AU and 3500AU. This
analysis therefore 
reveals that there is local order in the velocity field at these
scales. It is interesting that these scales are respectively disk
scales (140 AU) and outflow scales ($>$1000 AU), the latter
  being associated with (re-)injection of energy into the system, that
is, a driving scale. These scales are
approximately the same 
as those identified in the brightness data, which were found to be at 110 and 1700 AU, although there appears to be an
additional larger scale at 3500 AU in velocity. The weak structure in
the brightness at r$\sim$ 550AU and 
2700AU is however not seen in the velocity. An analysis of the
velocity data has also been performed without brightness weighting
with essentially the same result as in Fig.~\ref{fig:velocity}.

\subsection{Systematic errors associated with estimates of structure scale}
\label{errors}

There are systematic errors inherent in the present method. As already noted, derivatives of higher order SFs tend to
pick out scales which are too small by as much as 10-20\%. It turns
out that when we consider a distribution of clump sizes about some
mean value, this causes an overestimation of the mean clump
size as we discuss below. These two effects tend to cancel but to an extent which cannot be
quantified.

The detection of a preferred scale size implies that there is a range of sizes
around the derived value. In common with other methods, our
method does not give any measure of the extent of such a range. Here
we test the effect of introducing a distribution of clump sizes about
some mean preferred scale using artificial images. In order to
illustrate the behaviour of such systems, we assume that the
sizes of clumps are gaussian distributed
 about the mean. Using simulations similar to those
in Sect.~\ref{sec:method} we 
have created a number of maps with 200 randomly positioned clumps. Three mean sizes of 40, 60 and 80 pixels were used. We have found that when 
the half-width is smaller than $\sim$5\% of the mean value, the SFs are
indistinguishable from the original of only one clump size and the
minimum of the SF derivatives is found at the mean value. Increasing
the half-width beyond $\sim$5\% causes the minimum to move towards
larger scales. 
If the half-width of the distribution is 10\% (20\%) 25\% of the mean value the
minimum is found at 1.07 (1.25) 1.50 times the mean value. For example,
if the data used in Fig.~\ref{fig:fourier} consist of clumps
distributed as a gaussian with a halfwidth of 20\% of the mean value,
the detected value of 
1600AU corresponds to a true mean value of the
sizes of structures of $\sim$ 1300AU. This analysis only conveys the
sensitivity of 
the systematic error to the width of the distribution but does not
allow an estimation of the width from observations. The distribution of
clump sizes about some preferred scale is directly linked to the IMF
and it would be useful to develop methods which were capable of making
a reliable width estimate. 

\section{Conclusion}
In conclusion the technique presented here, whilst suffering from
some degree of uncertainty and imprecision to an extent common to all
methods, has proved valuable for analysis of images in our own
work. With the present and future advent of higher spatial resolution
maps in the radio region of the spectrum, probing the star-forming
scale, for example obtained with the Plateau de Bure interferometer, the
Submillimeter Array, CARMA and ALMA, this tool may prove useful to an
increasing community of observers.

\begin{acknowledgements}

MG and DF would like to acknowledge the support of the Instrument
Center for Danish 
Astrophysics (IDA), funded by the Danish Natural Science Research
Council and the Aarhus Centre
for Atomic Physics (ACAP), funded by the Danish Basic Research
Foundation.

\end{acknowledgements}

\bibliographystyle{aa}
\bibliography{bibliography}

\end{document}